\documentclass[prb,twocolumn,tbtags,preprintnumbers,nofootinbib,superscriptaddress,floatfix]{revtex4-2}

\usepackage{amsfonts}
\usepackage{amsmath}
\usepackage{amssymb}
\usepackage{graphics}
\usepackage{epsfig}
\usepackage{verbatim}
\usepackage{textcomp}
\usepackage[T1]{fontenc}
\usepackage{natbib}
\usepackage[colorlinks=true,urlcolor=blue,linkcolor=blue,citecolor=blue]{hyperref}
\usepackage{comment} 
\usepackage{tabularx}
\usepackage{xcolor}
\usepackage{braket} 
\usepackage{float}

\begin{document}
	
\title{Gorini-Kossakowski-Sudarshan-Lindblad equation in different bases:\\ application to driven-dissipative two- and multilevel systems}

\author{O.~A.~Ilinskaya$^\dag$}
\email{ilinskaya@ilt.kharkov.ua}
\affiliation{B.~Verkin Institute for Low Temperature Physics and Engineering of the National Academy of Sciences of Ukraine, Kharkiv 61103, Ukraine}
\author{O.~V.~Ivakhnenko$^\dag$}
\affiliation{B.~Verkin Institute for Low Temperature Physics and Engineering of the National Academy of Sciences of Ukraine, Kharkiv 61103, Ukraine}
\affiliation{Center for Quantum Computing, RIKEN, Wako, Saitama, 351-0198, Japan}
\author{A.~I.~Ryzhov}
\affiliation{B.~Verkin Institute for Low Temperature Physics and Engineering of the National Academy of Sciences of Ukraine, Kharkiv 61103, Ukraine}
\author{O.~Yu.~Kitsenko}
\affiliation{B.~Verkin Institute for Low Temperature Physics and Engineering of the National Academy of Sciences of Ukraine, Kharkiv 61103, Ukraine}
\affiliation{V.~N.~Karazin Kharkiv National University, Kharkiv 61022, Ukraine}
\author{S.~N.~Shevchenko}
\affiliation{B.~Verkin Institute for Low Temperature Physics and Engineering of the National Academy of Sciences of Ukraine, Kharkiv 61103, Ukraine}
\affiliation{Department of Mathematics, Kyiv School of Economics, 03113 Kyiv, Ukraine}	

\begin{abstract}
An open quantum system can be described by a master equation, of which one of the most popular is the Gorini–Kossakowski–Sudarshan–Lindblad (GKSL) equation. We revisit description of driven-dissipative quantum systems focusing on the appropriate choice of the system’s basis and the respective transformations.  We consider the GKSL equation in different bases and calculate the dynamics for a qubit and for a qu\textit{d}it. An appropriate choice of the basis is a fundamental problem for theoretical consideration of open quantum systems and provides an opportunity to obtain the desired evolution in practice.
\end{abstract}

\date{\today }
\maketitle
\def\thefootnote{\dag}\footnotetext{These authors contributed equally to this work.}

\section{Introduction}
\label{Sec:Introduction}
The Gorini-Kossakowski-Sudarshan-Lindblad (GKSL) equation~\cite{Gorini1976,Lindblad1976,Chruscinski2017}, also known as the
Lindblad equation, which celebrates its half-century anniversary this year, is the most common Markovian master equation~\cite{Breuer2007,manzano2020}. While in the work of Gorini, Kossakowski, and Sudarshan~\cite{Gorini1976} the master equation is derived for an $N$-level quantum system, Lindblad independently obtained the master equation as a consequence of the description of the general form of the quantum dynamical semigroup generator~\cite{Lindblad1976}, which is a generalization of the result of Ref.~\cite{Gorini1976}. Interestingly, the master equation was initially obtained for a two-level system by Kossakowski~\cite{Kossakowski1973}. Besides the fact that the GKSL equation is solved in various fields of quantum physics, fundamental questions related to Markovian master equations in the Lindblad form are still being studied these days~\cite{Pleasance2025}.

There are different forms of and approaches to
master equations for a time-dependent system Hamiltonian \cite{ivakhnenko2023}. If the
Hamiltonian changes with time adiabatically, one obtains the master equation
in the Lindblad form with the time-dependent Lindblad (or collapse) operators written in
the instantaneous energy eigenbasis~\cite{albash2012}. In the case of a strong
periodic driving, it is suggested to use the  Floquet
basis for the GKSL equation~\cite{Breuer2007}. Also, the superadiabatic basis is suggested as the one needed for performing the secular approximation when deriving a master equation for an adiabatically time-dependent Hamiltonian~\cite{kamleitner2013}. Importantly, the dissipator in
the Lindblad form can be used beyond the adiabatic regime~\cite{yamaguchi2017}. 
In Ref.~\cite{Gulacsi2025} the dissipator for a generally driven quantum system is written in the static eigenstate basis, and the corrections due to the time dependence of the Hamiltonian are taken into account, with an example of a generally driven qubit. Some studies derive the master equation beyond the Lindblad form~\cite{Bernazzani2025}.

The problem of choosing an appropriate basis appears each time the master
equation is considered~\cite{silveri2017}. Usually, original problems are
formulated in a \textit{diabatic basis} related to physically observable states. Various physically distinguishable bases are used for diverse quantum systems: charge states for charge qubits, current states for flux qubits, spin orientations for spin qubits, conduction or valence bands
for graphene, molecular states or lattice bands for ultracold molecules,
confining voltage for quantum phase transitions~\cite{ivakhnenko2023}. Then the collapse operators should be defined in a correct basis -- we will
call it the \textit{basis for computations}. All components of the GKSL equation should be transferred to this basis, while the resulting observables can afterwards be evaluated in any basis, for example, transformed back to the diabatic one~\cite{tayebirad2010,krzywda2020,krzywda2021,szabo2025,Bonifacio2020}. 

The next step in describing dynamics of quantum systems, after formulating
the problem and writing the Hamiltonian in the diabatic basis, is to write
down the master equation correctly. This implies transferring from the
diabatic basis to the basis for computations. When studying dynamics of
driven-dissipative systems, one can find diverse approaches in the literature.
Some authors ignore the difference between the diabatic basis and the one for collapse operators
and define them in the former, e.g., in Ref.~\cite{Novelli2015}. For time-dependent
problems, though, it is more natural to use the instantaneous 
eigenbasis~\cite{xu2014,yip2018}. For a periodically driven quantum system, the Floquet states are widely used~\cite{Son2009,Keliri2026,Meneses2026,Gu2023} and transitions among Floquet states are studied experimentally~\cite{Yen2026}. Different Floquet master equations can be useful in different limiting cases~\cite{Mickiewicz2026}. Dynamics of quantum states in different bases was
theoretically investigated in Refs.~\cite{wubs2005,silveri2012,silveri2013,silveri2015,danga2016,bose2026} and experimentally in  Refs.~\cite{zenesini2009,tayebirad2010}. Such facts bring us to the formulation of the questions related to the choice of bases for consideration of driven-dissipative qu\textit{b}%
its and qu\textit{d}its. Is it correct to define the collapse operators in a
diabatic basis? What is the procedure for introducing the correct
basis for computations for the GKSL equation? How to extend the consideration
of the GKSL equation from a qubit to a qu\textit{d}it? We devote this work to these issues.

The rest of the paper is organized as follows. The next Section introduces the
GKSL equation with the details in the Appendices~\ref{GSKL_Detail}-\ref{Sec:AppendixAdiabatic}. Section~\ref{Sec:Bases}
introduces different bases relevant to the master-equation approach, and some details describing the first superadiabatic basis are presented in Appendix~\ref{Appendix-superadiabatic}. In Sec.~\ref{Sec:Interferometry} we illustrate and compare the {dynamics and} stationary states
(interferograms) in several bases. Section~\ref{subsec:Lindblad_basis_general} is devoted to the bases and transformations in the description of driven-dissipative multilevel systems, with an example of dynamics for two coupled qubits.
Section~\ref{Sec:Conclusions} presents the conclusions.

{\section{Gorini–Kossakowski–Sudarshan-Lindblad equation}}
The GKSL equation for the density matrix $\rho$ of the system with the Hamiltonian $H_\text{S}(t)$ coupled to the environment (or bath, or reservoir) reads
\begin{equation}
	\dot{\rho}=-\frac{i}{\hbar}\left[ H_\text{S}(t),\rho \right] +\sum_{k}\left(
		L_{k}\rho L_{k}^{\dag}-\frac{1}{2}L_{k}^{\dag}L_{k}\rho -\frac{1}{2}\rho
		L_{k}^{\dag}L_{k}\right),  \label{Lindblad}
\end{equation}
where different channels of dissipation are described by the Lindblad operators $L_{k}$, which can be time-dependent~\cite{albash2012,Kitsenko2026}. 

There are two ways to derive the GKSL equation. The first way is the microscopic derivation which considers that the system is coupled to a dissipative environment and the environment degrees of freedom are traced out. This derivation looks simple for a time-independent system Hamiltonian and is reviewed in Appendix~\ref{GSKL_Detail}. There, the following assumptions are needed:
\begin{itemize}
	\item the separability (no correlations between the system and its environment at $t=0$);
	\item the Born approximation (a weak coupling between the system and environment implying an equilibrium state of the environment);
	\item the Markov approximation (the time scale of a decay of the environment correlation functions is much shorter than the smallest characteristic time of the system);
	\item the secular approximation, or rotating-wave approximation (RWA, all fast-rotating terms in the interaction picture can be neglected).
\end{itemize}
The accuracy of the weak-coupling approximation for a driven-dissipative qubit is studied in Ref.~\cite{Teixeira2021}.
For a qubit, the impact of the environment results in relaxation and dephasing, described by the Lindblad operators written in the instantaneous eigenbasis
\begin{equation}\label{relaxation-and-dephasing}
	L_{\mathrm{relax}}=\sqrt{\Gamma _{1}}\,\sigma_-, \quad 
	L_{\phi }=\sqrt{\Gamma _{\phi }/2}\,\sigma _{z},
\end{equation}
where $\sigma_-=(\sigma_x - i\sigma_y)/2$, $\Gamma_1$ and $\Gamma_\phi$ are the relaxation and dephasing rates, respectively. We also define the decoherence rate $\Gamma_2=\Gamma_1/2 + \Gamma_\phi$.

The second way to derive the GKSL equation is to use completely positive and trace-preserving maps. Such a derivation is reviewed in Appendix~\ref{Sec:AppendixGKSL-maps}. In Appendix~\ref{Sec:AppendixAdiabatic}, we (i)~note the difference between a transfer to the interaction representation for a time-independent and time-dependent system Hamiltonian, (ii)~give the adiabatic conditions for the Landau-Zener-St\"uckelberg-Majorana problem and a periodically driven qubit, and (iii)~following Ref.~\cite{Zvyagin1983}, make the qubit Hamiltonian time-independent for the small driving amplitude compared to the driving frequency.

\section{Different bases for quantum systems}\label{Sec:Bases}

\subsection{Importance of the basis choice in solving the master equation}
\begin{figure}[t]
	\includegraphics[width=8.4cm]{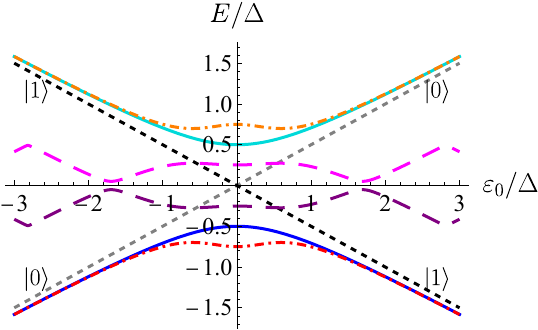}
\caption{The energy levels of a qubit as functions of the energy detuning $\varepsilon_0$ (both in the units of the energy gap $\Delta$) in different bases: the diabatic basis (the black and gray dashed lines), the instantaneous eigenbasis (the blue and cyan solid curves), the first superadiabatic basis (the red and orange dot-dashed curves), and the Floquet basis (the purple and magenta long-dashed curves). The labels $|0\rangle$ and $|1\rangle$ at the corners denote the diabatic eigenstates corresponding to the respective diabatic energy levels. The moment of time is fixed: $t_0=\pi/2\omega$, other parameters are as follows: the excitation amplitude $A=\Delta$, the driving frequency $\hbar\omega=\Delta$ (then, the parameter in Eq.~(\ref{parameter-a}) is $a=1$). For details, see Sec.~\ref{Sec:Bases}.}
	\label{Fig:energy-levels}
\end{figure}

In this section different bases are introduced for clarity on an example of a qubit (multilevel systems, qu\textit{d}its, are then considered in Sec.~\ref{subsec:Lindblad_basis_general}). The information about the bases is summarized in Table~\ref{table:bases} and illustrated in Fig.~\ref{Fig:energy-levels}. In what follows we consider diabatic, static, instantaneous, superadiabatic, and Floquet eigenstate bases (see Table~\ref{table:bases}). 

The qubit Hamiltonian in the diabatic basis (see Section~\ref{Sec:DiabaticBasis}) has the form of the pseudospin Hamiltonian
\begin{equation}\label{qubit-Hamiltonian}
	H(t)=-\frac{1}{2}\Big(\Delta\,\sigma_x+\varepsilon(t)\,\sigma_z\Big),
\end{equation}
where $\Delta$ is the energy gap and $\varepsilon(t)$ is a (time-dependent) bias. Below we consider 
\begin{equation}\label{harmonic-driving}
	\varepsilon(t)=\varepsilon_0+A\cos\omega t,
\end{equation}
where $\varepsilon_0$ is a constant detuning, while $A$ and $\omega$ are the amplitude and the (angular) frequency of the harmonic driving signal. Two distinct regimes of driving exist, namely, the adiabatic regime when the adiabaticity parameter 
\begin{equation}\label{delta}
	\delta=\frac{\Delta^2}{4 A\hbar\omega}
\end{equation}
is large and the opposite diabatic (or nonadiabatic) regime when $\delta$ is small.

An obvious but important note is that all operators in the master equation should be written in the same basis. Otherwise, an incorrect result would be obtained, for example, relaxation to the higher energy level. It should also be noted that a time-dependent Hamiltonian may lead to a time-dependent transfer matrix between bases.

\vspace{-0.6cm}

\subsection{Diabatic basis}\label{Sec:DiabaticBasis}
Consider so-called \textit{diabatic} basis. It is formed by the eigenstates of the operator 
$\sigma_z$, or, alternatively speaking, by the eigenstates of the qubit Hamiltonian~(\ref{qubit-Hamiltonian}) for the tunneling amplitude $\Delta=0$. We denote these eigenstates as follows
\begin{equation}\label{2-level-diabatic-basis}
	\ket{1}=\left(\begin{array}{cccc}
		1 \\
		0 \\
	\end{array}\right),\qquad
	\ket{0}=\left(\begin{array}{cccc}
		0 \\
		1 \\
	\end{array}\right).
\end{equation}
For example, for a flux qubit~\cite{Ilinskaya2024,Ilinskaya2026,Xiang2013}, these eigenvectors correspond to clockwise and counterclockwise directions of the current in the loop.
The corresponding eigenvalues of the Hamiltonian are $\mp\varepsilon(t)/2$. 

The Lindblad operators $L_\text{relax}$ and $L_\phi$ for a qubit, defined in Eq.~(\ref{relaxation-and-dephasing}), are not written  in the diabatic basis. Indeed, the diabatic state that corresponds to the lower energy on one side of the crossing point becomes the upper level on the other side of the crossing point. Therefore, for $\Gamma_1 > 0$ the operator $\sigma_-$ in this representation would incorrectly lead to energy relaxation on one side of the crossing point and to excitation on the other; the correct direction can be restored by using negative relaxation rates~\cite{Persson2010}. The Lindblad operators in Eq.~(\ref{relaxation-and-dephasing}) should then be written in the instantaneous eigenbasis (see Section~\ref{Subsec:Adiabatic}).
\begin{table*}[t]
	\centering
	\renewcommand{\arraystretch}{1.4} 
	\begin{tabular}{ | l | l | l | l | }
		\hline
		Basis name & Eigenstates of which operator & References & When used \\ \hline
		diabatic basis & $\sigma_z$ & \cite{Novelli2015} & diabatic regime (for interferograms) \\ \hline
		static eigenstate basis & $H_0$, Eq.~(\ref{H_0}) & \cite{Gulacsi2025} & adiabatic regime (for interferograms) \\ \hline
		instantaneous eigenbasis & $H(t)$, Eq.~(\ref{qubit-Hamiltonian}) & \cite{albash2012,xu2014, yip2018} & adiabatic regime (for dynamics) \\ \hline
		superadiabatic basis & $H_\text{ad}(t)$, Eq.~(\ref{Hamiltonian-adiabatic}) & \cite{kamleitner2013,GueryOdelin2019} & counterdiabatic driving \\
		\hline
		Floquet states basis & $H_\text{F}$, Eq.~\eqref{Floquet-Hamiltonian} & \cite{Son2009,Keliri2026,Meneses2026,Gu2023} & diabatic regime (for dynamics) \\
		\hline
	\end{tabular}
	\caption{ Different bases relevant in the context of the GKSL equation.}
	\label{table:bases}
\end{table*}

\subsection{Static eigenstate basis}
The \textit{static eigenstate} basis is formed by the eigenvectors of the static Hamiltonian
\begin{equation}\label{H_0}
	H_\text{0}=-\frac{1}{2}\left(\Delta\sigma_x+\varepsilon_0 \sigma_z\right),
\end{equation}
which is the Hamiltonian~(\ref{qubit-Hamiltonian}) with $\varepsilon(t)\equiv \varepsilon_0$. The eigenvectors are the following:
\begin{equation}\label{2-level-static-basis}
	\ket{E_-}=\left(\begin{array}{cccc}
		\gamma_+ \\
		\gamma_- \\
	\end{array}\right),\qquad
	\ket{E_+}=\left(\begin{array}{cccc}
		\gamma_- \\
		-\gamma_+ \\
	\end{array}\right),
\end{equation}
with the corresponding eigenvalues 
\begin{equation}
	E_{\pm}=\pm\frac{1}{2}\sqrt{\Delta^{2}+\varepsilon_0^{2}}=
	\pm\frac{1}{2}\Delta E \label{Epm}
\end{equation}
and 
\begin{equation}
	\gamma_\pm=\frac{1}{\sqrt{2}}
	\sqrt{1\pm\frac{\varepsilon_0}{\Delta E}}.
\end{equation}

For some applications 
that do not consider qubit level dynamics, the collapse operators can be written in the static eigenstate basis as it gives the correct direction of the dissipation on average. Static eigenstates are much simpler to deal with for numerical calculations. But the static eigenstate basis is unsuitable for describing the dynamics as it can also lead to the incorrect relaxation to the upper level, as does the diabatic basis. The reason is that the static eigenstate basis produces the correct direction of relaxation only on one side of the quasicrossing point.

\subsection{Instantaneous eigenbasis}\label{Subsec:Adiabatic}
\subsubsection{Instantaneous eigenstates}
The instantaneous eigenvalues $E_{\pm }(t)$ and eigenstates $\left\vert
E_{\pm }(t)\right\rangle $ are given by the Schr\"{o}dinger equation, where
the time is a parameter, $H(t)\left\vert E_{\pm }(t)\right\rangle =E_{\pm
}(t)\left\vert E_{\pm }(t)\right\rangle $. The \textit{instantaneous} eigenbasis (or \textit{adiabatic} basis) is formed by the eigenvectors of the Hamiltonian~(\ref{qubit-Hamiltonian}). The eigenenergies are as follows:
\begin{equation}
	{\color{black}E_{\pm }(t)=\pm \frac{1}{2}\sqrt{\Delta ^{2}+\varepsilon (t)^{2}}=\pm \frac{1%
	}{2}\Delta E(t).}\color{black}  \label{Epm(t)}
\end{equation}%
Here, $\Delta E(t)=E_{+}(t)-E_{-}(t)$ is the distance between the energy levels.
The eigenvectors read
\begin{equation}\label{2-level-adiabatic-basis}
	\ket{E_-(t)}=\left(\begin{array}{cccc}
		\gamma_+(t) \\
		\gamma_-(t) \\
	\end{array}\right),\qquad
	\ket{E_+(t)}=\left(\begin{array}{cccc}
		\gamma_-(t) \\
		-\gamma_+(t) \\
	\end{array}\right).
\end{equation}
Here 
\begin{equation}
	\gamma_\pm(t)=\frac{1}{\sqrt{2}}
	\sqrt{1\pm\frac{\varepsilon(t)}{\Delta E(t)}}.
\end{equation}
The most general form of the transfer matrix from the diabatic basis~(\ref{2-level-diabatic-basis}) to the instantaneous eigenbasis~(\ref{2-level-adiabatic-basis}) is as follows:
\begin{equation}\label{2-level-S-general}
	S_\text{gen}(t)=
	\left(\begin{array}{cccc}
		e^{i\varphi_1}\gamma_+(t) & -e^{i\varphi_2}\gamma_-(t) \\
		e^{i\varphi_1}\gamma_-(t) & e^{i\varphi_2}\gamma_+(t) \\
	\end{array}\right),
\end{equation}
with $\varphi_1$ and $\varphi_2$ being real constants. In Section~\ref{Subsec:Superadiabatic} we use the transfer matrix with $\varphi_1=0$ and $\varphi_2=\pi$, namely,
\begin{equation}\label{2-level-S}
	S(t)=
	\left(\begin{array}{cccc}
		\gamma_+(t) & \gamma_-(t) \\
		\gamma_-(t) & -\gamma_+(t) \\
	\end{array}\right).
\end{equation}
We note that the transfer matrix in Eq.~(\ref{2-level-S}) is equal to its inverse: 
$S_{\text{d}\to\text{ad}}(t)=S_{\text{ad}\to\text{d}}(t)\equiv S(t)$.

\subsubsection{Solution of the master equation, approach 1: transfer of the Liouville-von Neumann equation to the instantaneous eigenbasis}
Here we solve the GKSL equation with the collapse operators defined in the instantaneous eigenbasis. For this, we start from the Liouville-von Neumann equation, which is Eq.~(\ref{Lindblad}) taken without the dissipative terms, and transfer it to the instantaneous eigenbasis. 
The transfer from the diabatic to the instantaneous eigenbasis can alternatively be done using a transfer matrix, distinct from the matrix $S(t)$ in Eq.~(\ref{2-level-S}). Specifically, one can make a rotation around the $y$ axis using the matrix~\cite[p.~40]{Shevchenko2019}
\begin{equation}
	\tilde{S}(t)=e^{i\sigma_y \theta(t)/2},
\end{equation}
where 
\begin{equation}\label{tilde-theta}
	\theta(t)=-\arctan[\Delta/\varepsilon(t)].
\end{equation}
Then, the Hamiltonian in the instantaneous eigenbasis reads
\begin{align}\label{Hamiltonian-adiabatic-SectionD}
	&\tilde{H}_\text{ad}(t)=
	\tilde{S}^{-1}(t)H(t)
	\tilde{S}(t)
	-i\hbar\tilde{S}^{-1}(t)
	\dot{\tilde{S}}(t)\nonumber\\
	&\hspace{1cm} = -\frac{\Delta E(t)}{2}\sigma_z 
	+ \frac{\hbar\dot{\theta}(t)}{2}\sigma_y,
\end{align}
which includes the Berry connection term, the second term on the right-hand side, see Ref.~\cite{Berry1987}. The first term in the first line in Eq.~(\ref{Hamiltonian-adiabatic-SectionD}) is the diagonal matrix equal to the first term in the second line. In Eq.~(\ref{Hamiltonian-adiabatic-SectionD}) 
\begin{equation}\label{nonadiabatic-coupling-SectionD}
	\hbar\dot{\theta}(t)=-a
	\cdot\frac{\Delta^3\sin\omega t}{\Delta^2+\varepsilon^2(t)},
\end{equation}
{where 
	\begin{equation}\label{parameter-a}
		a=\frac{A\hbar\omega}{\Delta^2}.
	\end{equation}
The parameter $a$ is related to the adiabaticity parameter~(\ref{delta}) as follows: $a = 1/4\delta$, see Ref.~\cite{ivakhnenko2023}; the meaning of the adiabaticity parameter becomes clear if we consider the Landau-Zener-St\"uckelberg-Majorana (LZSM) probability of a nonadiabatic transition: $P_\text{LZSM} = \exp(-2\pi\delta)$. In our consideration here the parameter $a$ is more natural since it is small in the adiabatic limit.

\subsubsection{Solution of the master equation, approach 2: transfer of collapse operators to the diabatic basis}
The GKSL equation can also be solved in the diabatic basis. Then, the Hamiltonian is taken in this basis and the Lindblad operators defined in Eq.~(\ref{relaxation-and-dephasing}) should be transferred to this basis as follows:
\begin{eqnarray}
	&L_\text{relax}^\text{(d)}(t)=\tilde{S}(t)L_\text{relax}\tilde{S}^{-1}(t), \\
	&L_\phi^\text{(d)}(t)=\tilde{S}(t)L_\phi\tilde{S}^{-1}(t).
\end{eqnarray}
The Lindblad operators become time-dependent, and this approach can be computationally more expensive than the first approach described above [see the sum on the right-hand side of Eq.~(\ref{Lindblad})].

\subsection{Superadiabatic bases}\label{Subsec:Superadiabatic}
Since Berry's works~\cite{Berry1987,Berry1990,Lim1991}, the word `superadiabatic' is widely used in the literature: superadiabatic bases~\cite{Drese1998,kamleitner2013,Zhelnin2025}, superadiabatic regime~\cite{Lima2025}, {superadiabatic driving~\cite{Theisen2017,Giannelli2014}, superadiabatic population transfer~\cite{Vepsaelaeinen2019}}. The superadiabatic bases are used for counterdiabatic driving when the goal is to control the transition between levels~\cite{Theisen2017}.
In defining superadiabatic bases we follow Refs.~\cite{Berry1987,Drese1998}. However, unlike these studies, we do not artificially transfer to a `slow' time by using a small parameter. This small parameter appears naturally as shown below. For concreteness we consider the Hamiltonian~(\ref{qubit-Hamiltonian}) of a periodically driven qubit. The Schr\"odinger equation in the diabatic basis reads
\begin{equation}\label{Schrodinger-equation}
	i\hbar\frac{d\ket{\psi(t)}}{dt}=H(t)\ket{\psi(t)},
\end{equation}
with the Hamiltonian $H(t)$ defined in Eq.~(\ref{qubit-Hamiltonian}).
The Schr\"odinger equation in the instantaneous eigenbasis is as follows:
\begin{equation}\label{Schrodinger-adiabatic-basis}
	i\hbar \frac{d\ket{\psi_\text{ad}(t)}}{dt}=H_\text{ad}(t)\ket{\psi_\text{ad}(t)},
\end{equation}
where
\begin{align}\label{Hamiltonian-adiabatic}
	&H_\text{ad}(t)=
	S^{-1}(t)H(t)
	S(t)
	-i\hbar S^{-1}(t)
	\dot{S}(t)\nonumber\\
	&\hspace{1cm} = -\frac{\Delta E(t)}{2}\sigma_z 
	- \frac{\hbar\dot{\theta}(t)}{2}\sigma_y,
\end{align}
with $S(t)$ defined in Eq.~(\ref{2-level-S}).
The transfer matrix from the instantaneous basis to the first superadiabatic basis is built from the eigenvectors of the Hamiltonian~(\ref{Hamiltonian-adiabatic}) (see explanation below and Appendix~\ref{Appendix-superadiabatic}).
The first term in the first line on the right-hand side of Eq.~(\ref{Hamiltonian-adiabatic}) is a diagonal matrix $-\Delta E(t)\sigma_z/2$, while the second term appears due to the time dependence of the transfer matrix $S(t)$ and is called nonadiabatic coupling~\cite{Drese1998}.
We note that the nonadiabatic coupling obtained by using the transfer matrices $S(t)$ and $\tilde{S}(t)$ differs by a sign [compare Eqs.~(\ref{Hamiltonian-adiabatic-SectionD}) and (\ref{Hamiltonian-adiabatic})]. That illustrates the fact that the Hamiltonians in the superadiabatic bases are not uniquely defined since they depend on the form of the Hamiltonian in the instantaneous eigenbasis (see the explanation below and Appendix~\ref{Appendix-superadiabatic}).
If the Hamiltonian $H(t)$ changes adiabatically in time, the parameter $a$, Eq.~(\ref{parameter-a}),
is small [$a\ll 1$, see Eq.~(\ref{adiabaticity-parameter})]. 

A further unitary transformation, which approximately diagonalizes the Hamiltonian $H_\text{ad}(t)$, Eq.~(\ref{Hamiltonian-adiabatic}), leads to the Hamiltonian $H_1(t)$, Eq.~(\ref{Hamiltonian-1-superadiabatic}), in the first superadiabatic basis (see Appendix~\ref{Appendix-superadiabatic} for details). The energy levels in the first superadiabatic basis (as well as in the diabatic, instantaneous, and Floquet eigenbases) are illustrated in Fig.~\ref{Fig:energy-levels}. 

The difference between the energy levels in the first superadiabatic basis and those in the instantaneous eigenbasis is due to a time dependence of the Hamiltonian. This difference is given by the second term in the parentheses in front of $\sigma_z$ in Eq.~(\ref{Hamiltonian-1-superadiabatic}), and for $\varepsilon_0=0$ and $t_0=\pi/2\omega$ it is equal to $d=a^2\Delta/4$. The driving amplitude $A$ and frequency $\omega$ are chosen so that the parameter $a$ is not small $(a=1)$ for illustrative purposes. The moment of time $t_0$ in Fig.~\ref{Fig:energy-levels} is chosen so that $\dot{\theta}(t_0)\neq 0$. In the Hamiltonian~(\ref{Hamiltonian-1-superadiabatic}) the nonadiabatic coupling is of the order of $a^2$. 

The Hamiltonian in the second superadiabatic basis is obtained by the approximate diagonalization of the Hamiltonian in the first superadiabatic basis, Eq.~(\ref{Hamiltonian-1-superadiabatic}). Reiterating the procedure of subsequent diagonalization $n$ times, one obtains in the $n^\text{th}$ superadiabatic basis the nonadiabatic coupling of the order of $a^{n+1}$~\cite{Drese1998}.

Originally, the superadiabatic bases were used to analytically calculate the time dependence of a transition amplitude for a $2\times 2$ Hamiltonian~\cite{Lim1991}. There, it was shown that an optimal superadiabatic basis exists with the number $n=n_\text{opt}$, which in particular depends on the small parameter $a$. The transition amplitude calculated in the optimal superadiabatic basis has the best agreement with the numerical solution of the Schr\"odinger equation, while the agreement disappears for $n>n_\text{opt}$~\cite{Lim1991}.


\subsection{Floquet states basis}\label{Floquet}
For a periodically driven quantum system, the solution to the Schr\"odinger equation can be obtained by using the Floquet theorem~\cite{Breuer2007}. According to this theorem, Ref.~\cite{Son2009,Keliri2026,Meneses2026,Gu2023}, applied to a periodically driven qubit with the Hamiltonian~(\ref{qubit-Hamiltonian}), there exists a basis of solutions to the Schr\"odinger equation~(\ref{Schrodinger-equation}):
\begin{equation}\label{Floquet-basis}
	\ket{\Psi_j(t)}=e^{-i e_j t/\hbar}\ket{\Phi_j(t)},
\end{equation}
where $j=1,2$ for a qubit, the real constants $e_j$ are the quasienergies, and $\ket{\Phi_j(t+T)}=\ket{\Phi_j(t)}$, with $T=2\pi/\omega$ being the period of the driving signal in Eq.~(\ref{harmonic-driving}). The functions $\ket{\Phi_j(t)}$ are the Floquet states~\cite{Breuer2007}. Substituting the solutions~(\ref{Floquet-basis}) into the Schr\"odinger equation~(\ref{Schrodinger-equation}), one obtains the following equation:
\begin{equation}\label{differential-Floquet}
	\left[H(t)-i\hbar\frac{d}{dt}\right]\ket{\Phi_j(t)}=e_j \ket{\Phi_j(t)}.
\end{equation}
It can be easily verified that the function
\begin{equation}
	\ket{\Phi_j(t)} e^{i\omega m t} \quad (m\in\mathbb{Z})
\end{equation}
is also a solution to Eq.~(\ref{differential-Floquet}), but with a different quasienergy, specifically, $e_j + \hbar\omega m$.
The Floquet theorem allows one to transfer from this system of two differential equations to an infinite system of linear equations for the Fourier components of the Floquet states. Expanding the Floquet states $\ket{\Phi_j(t)}$ in a Fourier series:
\begin{equation}
	\ket{\Phi_j(t)}=\sum_{n=-\infty}^\infty \ket{\Phi_{j,n}}e^{i n \omega t},
\end{equation}
one finds the following system of equations:
\begin{align}\label{Floquet-system}
&\left(H_\text{F}^{[0]}+\hbar\omega n\right)\ket{\Phi_{j,n}} + H_\text{F}^{[1]}\ket{\Phi_{j,n-1}} + 
H_\text{F}^{[-1]}\ket{\Phi_{j,n+1}}\nonumber\\
&\hspace{1.5cm}=e_j \ket{\Phi_{j,n}},
\end{align}
where $H_\text{F}^{[0]}=H_0$ [see the static Hamiltonian in Eq.~(\ref{H_0})] and $H_\text{F}^{[1]}=H_\text{F}^{[-1]}=-A\sigma_z/4$. The left-hand side of the system of equations~(\ref{Floquet-system}) can be written as the following three-block-diagonal matrix:
\begin{widetext}
\begin{equation}\label{Floquet-Hamiltonian}
	H_\text{F}=
	\left(
	\begin{array}{c|cc|cc|cc|c}
		\ddots & & & & & & & \\
		\hline 
		&
		\vphantom{\Biggl(} -\dfrac{\varepsilon_0}{2}+(n-1)\hbar\omega \vphantom{\Biggr)} &
		-\dfrac{\Delta}{2} &
		-\dfrac{A}{4} &
		0 &
		0 &
		0 &
		\\
		&
		\vphantom{\Biggl(}-\dfrac{\Delta}{2} \vphantom{\Biggr)} &
		\dfrac{\varepsilon_0}{2}+(n-1)\hbar\omega &
		0 &
		\dfrac{A}{4} &
		0 &
		0 & 
		\\
		\hline
		&
		\vphantom{\Biggl(}-\dfrac{A}{4} \vphantom{\Biggr)} &
		0 &
		-\dfrac{\varepsilon_0}{2}+n\hbar\omega &
		-\dfrac{\Delta}{2} &
		-\dfrac{A}{4} &
		0 &
		\\
		&
		0 &
		\vphantom{\Biggl(} \dfrac{A}{4} \vphantom{\Biggr)} &
		-\dfrac{\Delta}{2} &
		\dfrac{\varepsilon_0}{2}+n\hbar\omega &
		0 &
		\dfrac{A}{4} &
		\\
		\hline
		&
		0 &
		0 &
		\vphantom{\Biggl(} -\dfrac{A}{4} \vphantom{\Biggr)} &
		0 &
		-\dfrac{\varepsilon_0}{2}+(n+1)\hbar\omega &
		-\dfrac{\Delta}{2} &
		\\
		&
		0 &
		0 &
		0 &
		\vphantom{\Biggl(} \dfrac{A}{4} \vphantom{\Biggr)} &
		-\dfrac{\Delta}{2} &
		\dfrac{\varepsilon_0}{2}+(n+1)\hbar\omega &
		\\
		\hline
		& & & & & & & \ddots 
	\end{array}
	\right),
\end{equation}
\end{widetext}
multiplied by the vector
\begin{equation}\left(\dots, \Phi_{1,n-1}, \Phi_{2,n-1}, \Phi_{1,n}, \Phi_{2,n}, \Phi_{1,n+1}, \Phi_{2,n+1}, \dots\right)^\intercal.
\end{equation} 
The system of equations~(\ref{Floquet-system}) can be solved numerically by truncating to a finite number of equations~\cite{Son2009}. {Alternatively, one can find an analytical solution when the energy gap $\Delta$ is considered as a small parameter: $\Delta\ll A,\hbar\omega$. Then, the first order in the perturbation theory in $\Delta$} leads to the result for the upper-level occupation probability obtained by the rotating-wave approximation~\cite{Shevchenko2019}, while the second order is the result of the generalized Van Vleck perturbation theory~\cite{Son2009}.
\begin{figure*}[t]
	\includegraphics[width=1.0 \textwidth]{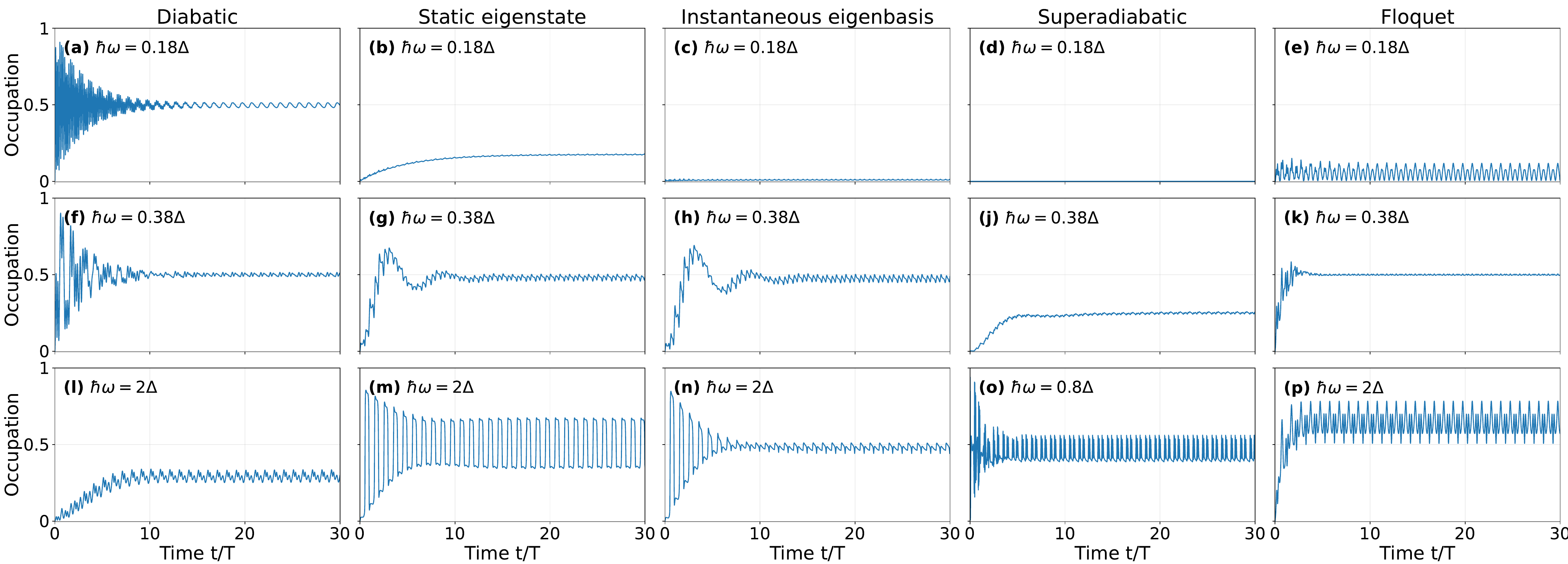}
	\centering \caption{Comparison of the dynamics of the upper energy level occupation probability for different driving frequencies $\hbar\omega=0.18\,\Delta$, $0.38\,\Delta$, $2\,\Delta$ (the three rows) in different bases (the five columns), with high relaxation and decoherence rates $\Gamma_1=0.15\Delta/\hbar$, $\Gamma_2=0.175\Delta/\hbar$. The horizontal axes show the time in periods of the driving signal $T=2\pi/\omega$.  The energy detuning is $\varepsilon_0=0.96\,\hbar\omega$, the driving amplitude is $A=4.02\,\hbar\omega$. We note a lower value of the driving frequency in panel~(o), $\hbar\omega=0.8\,\Delta$, compared to other panels in the third row. The reason is that the superadiabatic basis is inapplicable in the nonadiabatic regime. The parameter $a$, see Eq.~(\ref{parameter-a}), is approximately 0.13 for the first row, 0.58 for the second row, and 16.08 for the third row, except panel~(o), where $a\approx 2.57$.  
		The initial condition is as follows: in each basis (for every column), the ground state is occupied with the probability equal to unity, while other matrix elements of the density operator are equal to zero.
		The first column: (a,f,l) dynamics in the diabatic basis~(\ref{2-level-diabatic-basis}).  
		The second column: (b,g,m) dynamics in the static eigenstate basis~(\ref{2-level-static-basis}).
		The third column: (c,h,n) dynamics in the instantaneous eigenbasis, Eq.~(\ref{2-level-adiabatic-basis}).
		The fourth column: (d,j,o) dynamics in the first superadiabatic basis formed by the vectors~(\ref{superadiabatic-eigenvector-1}, \ref{superadiabatic-eigenvector-2}). These vectors are the approximate (to the first order in $a$) eigenvectors of the Hamiltonian~(\ref{Hamiltonian-adiabatic}). The fifth column: (e,k,p) dynamics calculated in the Floquet states basis, Eq.~(\ref{Floquet-basis}), and represented in the static eigenstate basis.
		}
	\label{Fig:Different Bases Dynamics comparison}
\end{figure*}

\section{Illustration: Dynamics and interferometry for a qubit in different bases}
\label{Sec:Interferometry}
\begin{figure*}[t]
	\includegraphics[width=1.0 \textwidth]{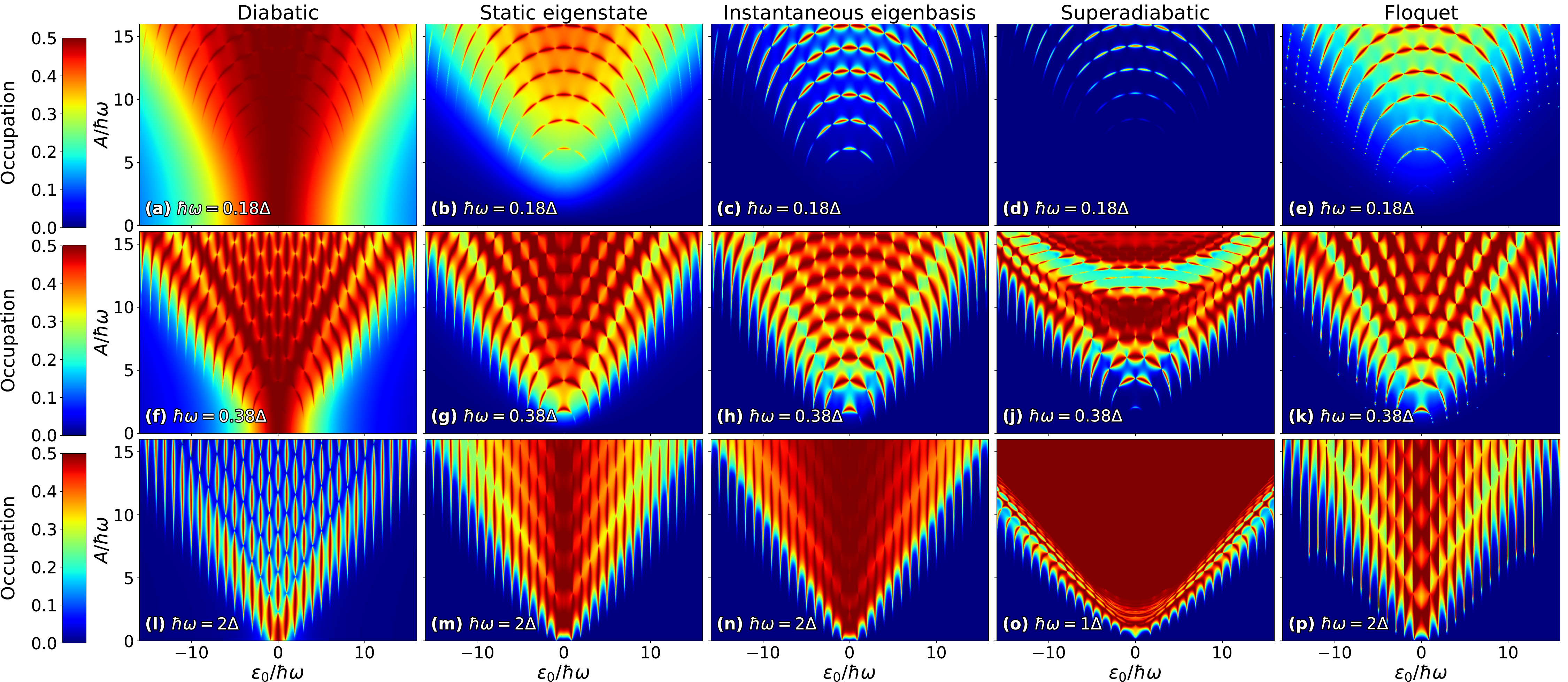}
	\centering \caption{Comparison of the interferograms (the time-averaged upper-level qubit occupation probability as a function of the energy detuning $\varepsilon_0$ and the driving amplitude $A$) for different driving frequencies $\hbar\omega=0.18\,\Delta$, $0.38\,\Delta$, $2\,\Delta$ (the three rows) in different bases (the five columns), with moderate relaxation and decoherence rates $\Gamma_1=0.03\Delta/\hbar$, $\Gamma_2=0.035\Delta/\hbar$. We note a lower value of the driving frequency in panel~(o), $\hbar\omega=\Delta$, compared to other panels in the third row. The reason is that the superadiabatic basis is inapplicable in the nonadiabatic regime. 
		The first column: (a,f,l) the interferograms in the diabatic basis~(\ref{2-level-diabatic-basis}), which show the cleanest results in the case of fast nonadiabatic driving.  
		The second column: (b,g,m) the interferograms in the static eigenstate basis~(\ref{2-level-static-basis}). The resonances are blurred for all driving frequencies. {Unlike the interferograms in the diabatic basis, these interferograms do not show any resonances at the point ($A=0$, $\varepsilon_0=0$). The reason is that the levels exhibit quasicrossing instead of crossing.}  
		The third column: (c,h,n) the interferograms in the instantaneous eigenbasis~(\ref{2-level-adiabatic-basis}). They show the cleanest results in the adiabatic regime (with $a\ll 1$).
		The fourth column: (d,j,o) the interferograms in the first superadiabatic basis formed by the vectors~(\ref{superadiabatic-eigenvector-1}, \ref{superadiabatic-eigenvector-2}). These vectors are the approximate (to the first order of $a$) eigenvectors of the Hamiltonian~(\ref{Hamiltonian-adiabatic}). As for large amplitudes $A$ and large driving frequencies $\omega$ the parameter $a$ is not small, the superadiabatic basis fails in the nonadiabatic regime and can only be applied in the deep adiabatic driving regime. The fifth column: (e,k,p) interferograms calculated in the Floquet states basis, Eq.~(\ref{Floquet-basis}), and represented in the static eigenstate basis.
		}
	\label{Fig:Different Bases Interferograms comparison}
\end{figure*}
This section illustrates different bases considered in the previous section for a periodically driven qubit, namely, the dynamics and interferograms are shown in Figs.~\ref{Fig:Different Bases Dynamics comparison} and~\ref{Fig:Different Bases Interferograms comparison}, respectively. The dynamics in different bases (the upper-level qubit occupation probabilities as functions of the time in units of the period $T=2\pi/\omega$ of the driving signal) is shown in Fig.~\ref{Fig:Different Bases Dynamics comparison}. In Fig.~\ref{Fig:Different Bases Dynamics comparison}(a,f,l), corresponding to the diabatic basis, the dynamics of the occupation probability shows the cleanest result in the case of fast (nonadiabatic) driving. For the static eigenstate basis, Fig.~\ref{Fig:Different Bases Dynamics comparison}(b,g,m), the decoherence can have a wrong direction on one side of the quasicrossing point because of time-independent Lindblad operators. That leads to a slight excitation every half-period and to a halo around the resonance lines in Fig.~\ref{Fig:Different Bases Interferograms comparison}(b) below. In Fig.~\ref{Fig:Different Bases Dynamics comparison}(c,h,n), corresponding to the instantaneous eigenbasis, the cleanest results are in the adiabatic regime (with $a\ll 1$).
The superadiabatic basis, see Fig.~\ref{Fig:Different Bases Dynamics comparison}(d,j,o), fails in the nonadiabatic regime (for large amplitudes $A$ and large driving frequencies $\omega$ when the parameter $a$ is not small) and can only be applied in the deep adiabatic driving regime. In Fig.~\ref{Fig:Different Bases Dynamics comparison}(e,k,p) corresponding to the Floquet basis, the dynamics was obtained by the numerical calculation of the single-period evolution  propagator applied consecutively to obtain the multi-period evolution. The collapse operators used in the Floquet basis were $\sigma_x$ for relaxation and $\sigma_z$ for dephasing~\cite{Johansson2013}.
In all panels of Fig.~\ref{Fig:Different Bases Dynamics comparison} the collapse operators and the initial conditions were defined in the basis, in which the result was displayed. We note the small and large periods in Fig.~\ref{Fig:Different Bases Dynamics comparison}(g,h), corresponding to the driving and Rabi periods, respectively.

In Fig.~\ref{Fig:Different Bases Interferograms comparison} we show that the same input parameters for interferometry {give different} resonance patterns in different bases. The time-averaging is done over 240 driving periods, which is enough to cover many Rabi-oscillation periods, if they are present, to get the correct average value. The Rabi oscillations decay, see Fig.~\ref{Fig:Different Bases Dynamics comparison}(g,h), and only the driving frequency impacts the time-averaged occupation probabilities. The maximum possible value of the time-averaged upper-level qubit occupation probability is 0.5 as the population inversion is not possible provided the collapse operators are correctly defined. We note that adiabatic bases ({static eigenstate, instantaneous}, superadiabatic) show the cleanest interferometry resonances in the adiabatic limit {$(a\ll 1)$}, while the diabatic basis provides the cleanest result in the diabatic limit {$(a\gg 1)$}. The superadiabatic basis fails rapidly when the driving becomes nonadiabatic at high driving amplitudes and driving frequencies, {$a\gtrsim 1$}. The Floquet basis interferometry in Fig.~\ref{Fig:Different Bases Interferograms comparison} was built with the QuTiP package \cite{qutip5}. For this fifth column, the steady state was calculated in the Floquet basis, Eq.~(\ref{Floquet-basis}), and the result for the averaged upper-level occupation probability was transferred back to the static eigenstate basis and displayed in that basis. \color{black}

\section{GKSL equation and basis choices for a multilevel quantum system}
\label{subsec:Lindblad_basis_general}

Consider generalization of the above treatment to the case of multilevel systems. Common examples of multilevel systems are two or more coupled qubits: flux qubits~\cite{Ploeg2007}, transmons~\cite{Kubo2024}, fluxonium qubits~\cite{Zhang2024}. Two- and multilevel systems can also be based on trapped ions~\cite{Randall2018}. We stress again that before substituting operators into the GKSL equation, 
it is necessary to express the Hamiltonian, density operator, and all Lindblad operators 
in a \emph{single, consistent basis}.  
Different physical processes are often most naturally expressed in different bases 
(e.g., diabatic, static eigenstate, or instantaneous eigenbasis), and transforming between them 
requires a careful definition of the unitary transformations. In this section we write general formulas without being tied to a specific Hamiltonian of a particular multilevel system and then provide an illustration: the dynamics for the two coupled qubits. All operators in this section are assumed to be time-dependent,
and all bases are orthonormal.

\paragraph{Diabatic basis.}
Let $\{\ket{n}\}$ denote a diabatic basis, typically corresponding to
well-defined charge, spin, or site configurations of the system.  
The microscopic Hamiltonian $H$ may generally include tunneling or coupling terms
between these basis states:
\begin{equation}
	H = \sum_{m,n} H_{mn} \ket{m}\bra{n}.
\end{equation}
This basis is often used when the coupling to external reservoirs, drives, or measurement devices 
is diagonal or has a simple form in terms of these physical degrees of freedom.
For instance, charge-tunneling, relaxation, and dephasing processes
may couple directly to the populations or coherences in this basis.

\paragraph{Natural (unsorted) instantaneous eigenbasis.}
The instantaneous eigenbasis of the system is defined by
\begin{equation}
	H(\lambda) \ket{\psi_j(\lambda)} = E_j(\lambda) \ket{\psi_j(\lambda)},
\end{equation}
where $\lambda$ denotes a control parameter such as a detuning, gate voltage, or magnetic field.
Let $U_{\mathrm{unsort}}(\lambda)$ be the unitary matrix whose columns are the eigenvectors
$\ket{\psi_j(\lambda)}$.
Then the Hamiltonian is transformed as
\begin{align}
	&U_{\mathrm{unsort}}^\dagger(\lambda)\,H(\lambda)\,U_{\mathrm{unsort}}(\lambda)
	\nonumber\\
	&\hspace{1cm}= \mathrm{diag}(E_1(\lambda),E_2(\lambda),\dots,E_d(\lambda)),
\end{align}
where $d$ is the number of energy levels of the qu\emph{d}it.
Each eigenstate $\ket{\psi_j(\lambda)}$ is associated with a definite set of quantum numbers or symmetries (if applicable).
Operators that represent dissipative transitions between eigenstates, such as
relaxation operators~\cite{Gegg2016}
\begin{equation}
	L_{j\to k}(\lambda) \propto \ket{\psi_k(\lambda)}\bra{\psi_j(\lambda)} \quad (j>k)
\end{equation}
and dephasing operators~\cite{Gegg2016}
\begin{equation}
	L_{\phi,jk}(\lambda)\propto\ket{\psi_j(\lambda)}\bra{\psi_j(\lambda)}-\ket{\psi_k(\lambda)}\bra{\psi_k(\lambda)} \quad (j>k),
\end{equation}
are naturally defined in this basis.

\paragraph{Energy-sorted instantaneous eigenbasis.}
It is sometimes convenient to relabel the eigenstates according to the instantaneous energy order~\cite{Gegg2016},
\[
E_1(\lambda) \le E_2(\lambda) \le \cdots \le E_d(\lambda),
\]
defining an \emph{energy-sorted instantaneous eigenbasis}
$\{\ket{E_1(\lambda)}, \ket{E_2(\lambda)}, \dots, \ket{E_d(\lambda)}\}$.
In this representation, the Hamiltonian remains diagonal:
\begin{equation}
	U_{\mathrm{sort}}^\dagger(\lambda)\,H(\lambda)\,U_{\mathrm{sort}}(\lambda)
	= \mathrm{diag}(E_1,\dots,E_d),
\end{equation}
where $U_{\mathrm{sort}}(\lambda) = U_{\mathrm{unsort}}(\lambda)\,\Pi(\lambda)$ and
$\Pi(\lambda)$ is a permutation matrix that ensures ascending energy order.
If the energy levels cross as a function of~$\lambda$ (see, for example, Ref.~\cite{Chatterjee2018}),
the permutation matrix changes discontinuously,
dividing the parameter space into regions where the ordering differs.
This representation is particularly useful when discussing adiabatic evolution,
Landau-Zener-St\"uckelberg-Majorana transitions, or thermodynamic properties,
though it introduces formal discontinuities in~$\lambda$~\cite{ivakhnenko2023}.

\paragraph{Choice of working basis.}
When the GKSL equation is formulated in the diabatic basis,
each Lindblad operator defined in the eigenbasis can be transformed as
\begin{equation}
	L^{\text{(d)}}_{\alpha} = U_{\mathrm{inst}}\, L_{\alpha}\, U_{\mathrm{inst}}^\dagger,
\end{equation}
where $U_{\mathrm{inst}} \in \{U_{\mathrm{unsort}}, U_{\mathrm{sort}}\}$ is the transfer matrix between diabatic basis and either the natural (unsorted) or the sorted instantaneous eigenbasis.
The dissipator retains its Lindblad-type form~\cite{manzano2020}:
\begin{equation}
	\breve{L}^{\text{(d)}}_{\alpha}[\rho^{\text{(d)}}]
	= L^{\text{(d)}}_{\alpha} \rho^{\text{(d)}} L^{\text{(d)}\dagger}_{\alpha}
	- \tfrac{1}{2}\{L^{\text{(d)}\dagger}_{\alpha} L^{\text{(d)}}_{\alpha},\, \rho^{\text{(d)}}\}.
	\label{eq:dissipator_general}
\end{equation}
The corresponding master equation reads
\begin{equation}
	\frac{d\rho^{\text{(d)}}}{dt}
	= -\frac{i}{\hbar}[H^{\text{(d)}},\rho^{\text{(d)}}]
	+ \sum_{\alpha}\breve{L}^{\text{(d)}}_{\alpha}[\rho^{\text{(d)}}].
	\label{eq:Lindblad_diabatic_general}
\end{equation}
Here, $H^{\text{(d)}}$ and $\rho^{\text{(d)}}$ denote operators in the diabatic basis.

Alternatively, when the GKSL equation is written in the instantaneous eigenbasis (either unsorted or sorted),
\begin{align}
	&\frac{d\rho}{dt}
	= -\frac{i}{\hbar}\Bigl[U_{\mathrm{inst}}^\dagger H^{\text{(d)}} U_{\mathrm{inst}}
	- i\hbar\,U_{\mathrm{inst}}^\dagger \frac{\partial U_{\mathrm{inst}}}{\partial t},\, \rho\Bigr]\nonumber\\
	&\hspace{1.1cm}+ \sum_{\alpha}\breve{L}_{\alpha}[\rho],
	\label{eq:Lindblad_eigenbasis_general}
\end{align}
there is an additional Berry connection term (the second term in the brackets, see also Eq.~(\ref{Hamiltonian-adiabatic}) and the explanation below it).
The density matrices are related by $\rho^\text{(d)} = U_{\mathrm{inst}}\,\rho\,U_{\mathrm{inst}}^\dagger$.
If the eigenstates vary smoothly with~$\lambda(t)$, the natural eigenbasis provides a continuous and numerically stable representation, whereas energy sorting may lead to artificial discontinuities at crossings.

We illustrate the dynamics of a multilevel system in different bases on an example of two coupled qubits. We consider the following Hamiltonian~\cite{Ploeg2007}:
\begin{align}\label{4-level-Hamiltonian}
	&H(t)=-\frac{\Delta_1}{2}\sigma_x\otimes {\hat 1}
	-\frac{\varepsilon_1(t)}{2}\sigma_z\otimes {\hat 1}\nonumber\\
	&\hspace{1.1cm}-\frac{\Delta_2}{2}{\hat 1}\otimes\sigma_x 
	-\frac{\varepsilon_2(t)}{2}{\hat 1}\otimes\sigma_z
	+\frac{J}{2}\sigma_z\otimes\sigma_z,
\end{align}
written in the diabatic basis $(1,0,0,0)^\intercal$, $(0,1,0,0)^\intercal$, $(0,0,1,0)^\intercal$, $(0,0,0,1)^\intercal$. Here, $\sigma_{x,z}$ are the Pauli matrices, $\varepsilon_{j}(t)$ and $\Delta_j$ are the energy bias and energy gap for the $j^\text{th}$ qubit $(j=1, 2)$, $J$ is the interaction energy between the qubits. We assume the driving to be identical for both qubits~\cite{Gramajo2018,Ilinskaya2026}:
\begin{equation}
	\varepsilon_{j}(t)=\varepsilon_{0j}+A\sin(\omega t),
\end{equation}
where $\varepsilon_{0j}$ is the time-independent energy bias for the $j^\text{th}$ qubit, $A$ and $\omega$ are the driving amplitude and angular frequency, respectively. {We note that the GKSL equation for a multilevel system can be solved analytically under certain assumptions~\cite{Zvyagin2020,Zvyagin2026}.}
\begin{figure}
	\includegraphics[width=\columnwidth]{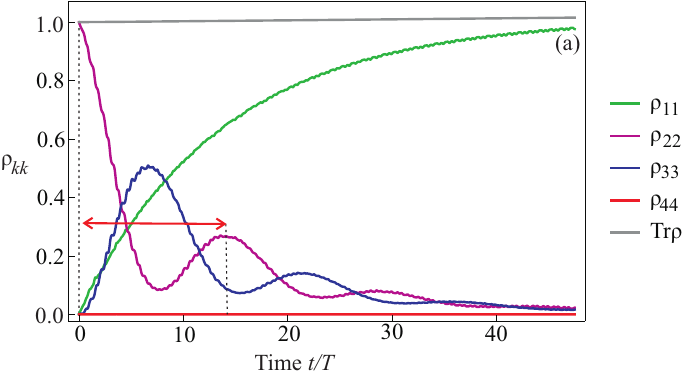}
	
	\vspace{0.2cm}
	\includegraphics[width=\columnwidth]{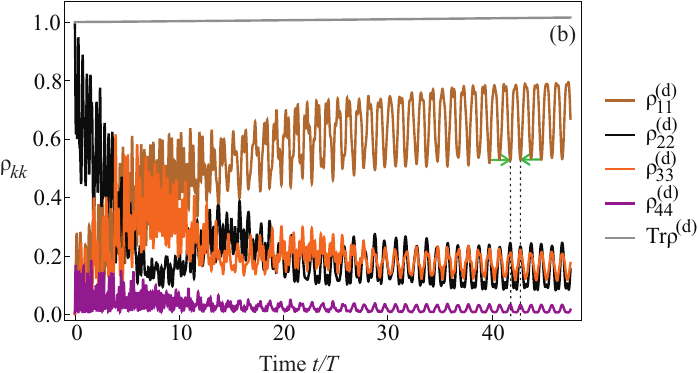}
	\caption{
		Dynamics for two driven qubits in different bases with relaxation and decoherence taken into account. The abscissa axes show time in units of the driving period $T=2\pi/\omega$. The driving frequency is resonant with the two eigenstates at $t=0$: $\hbar\omega=E_3(0)-E_2(0)$.  (a)~Dynamics in the instantaneous basis. The initial condition is as follows: $\rho_{22}(0)=1$, with other matrix elements being equal to zero.  (b)~Dynamics in the diabatic basis. The initial condition is as follows: $\rho_{22}^\text{(d)}(0)=1$, with other matrix elements being equal to zero. Both panels show Rabi-like oscillations. The driving period and the Rabi period are the same for both panels, and the Rabi period is shown by the red double arrow in panel~(a), while the driving period is shown by two green arrows in panel~(b). The simulation parameters are as follows: $\varepsilon_{01}=2\,\Delta_2$, $\varepsilon_{02}=1.7\,\Delta_2$, $A=0.5\,\Delta_2$, $J=0.82\,\Delta_2$, $\Delta_1=1.5\,\Delta_2$, $\Gamma=0.007\,\Delta_2/\hbar$, $\Gamma_\phi=0.0035\,\Delta_2/\hbar$.
		}
	\label{Fig:TwoQubits}
\end{figure}

We assume that the qubits are coupled to different baths~\cite{Li2008}. Then, the relaxation and dephasing superoperators in the instantaneous basis are as follows:
\begin{equation}
	\breve{L}_\text{relax}[\rho]=
	\frac{\Gamma}{2}\Big[2\sigma_{-}^{(j)}\rho(t)\sigma_{+}^{(j)}-
	\sigma_{+}^{(j)}\sigma_{-}^{(j)}\rho(t)-
	\rho(t)\sigma_{+}^{(j)}\sigma_{-}^{(j)}\Big]
\end{equation}
and
\begin{equation}
	\breve{L}_\phi[\rho]=
	\frac{\Gamma_{\phi}}{2}\Big[2\sigma_z^{(j)}\rho(t)\sigma_z^{(j)}-\rho(t)\Big],
\end{equation}
where 
$\sigma_\pm^{(1)}=\sigma_\pm\otimes\hat{1}$, $\sigma_\pm^{(2)}=\hat{1}\otimes\sigma_\pm$,
$\sigma_z^{(1)}=\sigma_z\otimes\hat{1}$, $\sigma_z^{(2)}=\hat{1}\otimes\sigma_z$, and
$\sigma_\pm=(\sigma_x\pm i\sigma_y)/2$. The relaxation and dephasing rates are assumed to be equal for both qubits and are denoted by $\Gamma$ and $\Gamma_{\phi}$. For the energy-level diagram of the two-qubit system, see Fig.~2(a) in Ref.~\cite{Ilinskaya2026}.
The dynamics for the two-qubit system in the instantaneous and diabatic bases is shown in Figs.~\ref{Fig:TwoQubits}(a) and \ref{Fig:TwoQubits}(b), respectively.

\section{Conclusions}
\label{Sec:Conclusions}
We revisited the description of driven-dissipative quantum systems focusing on the appropriate choice of the system's basis and the respective
transformations. Approximations used in the derivation of the GKSL equation
are the separability, the Born-Markov approximation, and the rotating-wave
approximation. We reviewed
different bases for quantum systems (diabatic, static eigenstate, instantaneous,
superadiabatic, and Floquet bases) and demonstrated the {dynamics and}
interferometry for a two-level system in these bases. It was
illustrated that the stationary states
(interferograms) have a qualitatively different appearance in different bases
and for various parameter choices. {We discussed the basis choice for a multilevel quantum system and gave an illustrative example of the dynamics for a four-level system.} Our consideration is potentially
important for theoretical calculations, where the understanding and choice of
diabatic basis and basis for computations is the fundamental problem, as well as for
quantum engineering, where switching between the representations may help in
obtaining desired evolution.

\section*{Acknowledgements}
The authors are grateful to B.~A.~Kushnarov 
for fruitful discussions and the referee for useful comments. O.A.I., O.Yu.K., and S.N.S. acknowledge financial support of the National Research Foundation of Ukraine (Grant No. 2025.07/0044). The work is partially supported by the U.S. National Academy of Sciences (NAS) and the Office of Naval Research (ONR) in the framework of the IMPRESS-U project, and NASU projects F16-15 and F26-5.

\appendix

\section{Microscopic derivation of the Gorini-Kossakowski-Sudarshan-Lindblad equation for a time-independent Hamiltonian}
\label{GSKL_Detail}
Here we consider the microscopic derivation of the GKSL equation following Ref.~\cite{Breuer2007}.
The full Hamiltonian in the Schr\"odinger representation 
\begin{equation}
	H_\text{T}=H_\text{S} + H_\text{B} + H_\text{I}
\end{equation}
consists of the system Hamiltonian $H_\text{S}$, the bath Hamiltonian $H_\text{B}$, and the Hamiltonian of the system--bath interaction $H_\text{I}$, which are all assumed to be time-independent. An operator $O(t)$ in the interaction representation is expressed through the corresponding operator $O$ in the Schr\"odinger representation as follows ($\hbar=1$ in this Appendix):
\begin{equation}\label{interaction-representation}
	O(t)=\exp\big(i(H_\text{S}+H_\text{B})t\big) O \exp\big(-i(H_\text{S}+H_\text{B})t\big).
\end{equation}
Below in this Appendix operators with a time dependence shown explicitly are written in the interaction representation. Starting from the Liouville-von Neumann equation for the whole system in the interaction representation
\begin{equation}
	\frac{d}{dt}\rho(t)=-i[H_\text{I}(t), \rho(t)]
\end{equation}
and using the separability condition and the Born-Markov approximation, we obtain after some labor~\cite{Breuer2007}
\begin{equation}\label{double-commutator}
	\frac{d}{dt}\rho_\text{S}(t)=-\int_0^\infty d\tau\, \text{tr}_\text{B} 
	\big[H_\text{I}(t),[H_\text{I}(t-\tau),\rho_\text{S}(t)\otimes\rho_\text{B}]\big].
\end{equation}
Here $\rho_\text{S}(t)$ is the reduced density operator of the system, $\rho_\text{B}$ is an equilibrium density operator of the bath, and $\text{tr}_\text{B}$ means tracing over the bath degrees of freedom.

It can be shown that the interaction Hamiltonian $H_\text{I}$ can be written as~\cite{Breuer2007}
\begin{equation}\label{decomposition-of-interaction}
	H_\text{I}=\sum_{\alpha,\omega} A_\alpha (\omega)\otimes B_\alpha = 
	\sum_{\alpha,\omega} A^\dag_\alpha (\omega)\otimes B^\dag_\alpha,
\end{equation}
where the operators $A_\alpha (\omega)$ and $B_\alpha$ act in the system and bath subspaces, respectively. Here $\omega=\varepsilon'-\varepsilon$ and $\varepsilon'$, $\varepsilon$ are the eigenvalues of the system Hamiltonian $H_\text{S}$. The operators $A_\alpha (\omega)$, and $A^\dag_\alpha (\omega)$ are the eigenoperators of $H_\text{S}$ with eigenvalues $-\omega$ and $\omega$, respectively, which means that
\begin{equation}\label{eigenoperators}
	[H_\text{S}, A_\alpha(\omega)]=-\omega A_\alpha(\omega), \quad 
	[H_\text{S}, A^\dag_\alpha(\omega)]=\omega A^\dag_\alpha(\omega).
\end{equation}
Equation~(\ref{eigenoperators}) implies that in the interaction representation we have
\begin{equation}
	A_\alpha(\omega,t)=e^{-i\omega t}A_\alpha(\omega).
\end{equation}
Transferring Eq.~(\ref{decomposition-of-interaction}) to the interaction representation and substituting it into Eq.~(\ref{double-commutator}), we obtain after some calculations~\cite{Breuer2007}
\begin{align}\label{before-RWA}
	&\frac{d}{dt}\rho_\text{S}(t)=\sum_{\omega,\omega'}\sum_{\alpha,\beta}
	e^{i(\omega'-\omega) t}\,\Gamma_{\alpha\beta}(\omega)\times\nonumber\\
	&\Big[A_\beta(\omega)\rho_\text{S}(t)A^\dag_\alpha(\omega')-
	A^\dag_\alpha(\omega')A_\beta(\omega)\rho_\text{S}(t)\Big]+\text{h.c.}.
\end{align}
Here, 
\begin{equation}\label{Fourier-transform}
	\Gamma_{\alpha\beta}(\omega)=\int_{0}^{\infty}ds\, e^{i\omega s}
	\langle B^\dag_\alpha(t)B_\beta(t-s)\rangle
\end{equation}
is a one-sided Fourier transform of the reservoir correlation function 
\begin{equation}\label{correlation-functions}
	\langle B^\dag_\alpha(t)B_\beta(t-s)\rangle=
	\langle B^\dag_\alpha(s)B_\beta(0)\rangle.
\end{equation}
Equation~(\ref{correlation-functions}) means that the reservoir correlation functions are time-independent and holds when the reservoir's state is stationary, i.e., $[\rho_\text{B}, H_\text{B}]=0$. 

If $\tau_\text{R}\gg\tau_\text{S}$, where $\tau_\text{R}$ is the relaxation time of the open system (equal to the inverse relaxation rate $\Gamma_1^{-1}$ for a qubit) and $\tau_\text{S}\sim|\omega'-\omega|^{-1}$ is a typical time scale of the system's own evolution, RWA can be applied. RWA means neglecting all the terms in Eq.~(\ref{before-RWA}) for which $\omega'\neq~\omega$. 

The matrix $\Gamma_{\alpha\beta}(\omega)$ in Eq.~(\ref{Fourier-transform}) can be written as
\begin{equation}\label{Gamma-decomposition}
	\Gamma_{\alpha\beta}(\omega)=\frac{1}{2}\gamma_{\alpha\beta}(\omega) +i 
	S_{\alpha\beta}(\omega),
\end{equation}
where 
\begin{equation}\label{matrix-S}
	S_{\alpha\beta}(\omega)=
	\frac{\Gamma_{\alpha\beta}(\omega)-\Gamma^\ast_{\beta\alpha}(\omega)}{2i}
\end{equation}
is a Hermitian matrix and 
\begin{align}\label{matrix-gamma}
	&\gamma_{\alpha\beta}(\omega)=
	\Gamma_{\alpha\beta}(\omega)+\Gamma^\ast_{\beta\alpha}(\omega)\nonumber\\
	&\hspace{1.1cm}=\int_{-\infty}^{\infty}ds\, e^{i\omega s}
	\langle B^\dag_\alpha(s)B_\beta(0)\rangle
\end{align}
is a Hermitian positive semi-definite matrix. 
	
While the hermiticity of the matrices in Eqs.~(\ref{matrix-S}, \ref{matrix-gamma}) can be simply shown, the proof that the matrix $\gamma_{\alpha\beta}(\omega)$ is positive semi-definite requires some calculations. One needs to show that
\begin{equation}\label{nonnegative-sum}
	\sum_{\alpha,\beta} v_\alpha^\ast \gamma_{\alpha\beta}(\omega) v_\beta \geq 0 \quad
	\forall v_{\alpha}\in \mathbb{C}.
\end{equation}
We define the reservoir operator as $X=\sum_\beta v_\beta B_\beta$, then, the expression in Eq.~(\ref{nonnegative-sum}) can be rewritten as follows:
\begin{equation}\label{nonnegative-sum-2}
	\sum_{\alpha,\beta} v_\alpha^\ast \gamma_{\alpha\beta}(\omega) v_\beta =
	\int_{-\infty}^{\infty}ds\, e^{i\omega s}\,
	\text{Tr}[\rho_\text{B}X^\dag(s)X].
\end{equation}
If the reservoir's state is stationary [see the text below Eq.~(\ref{correlation-functions})], the operators $\rho_\text{B}$ and $H_\text{B}$ can be measured simultaneously and there is a basis ${\ket{n}}$, in which they are both diagonal. In this basis $\rho_\text{B}$ has the form: $\rho_\text{B}=\sum_n p_n \ket{n}\bra{n}$, where $p_n$ are the probabilities. Then, the trace in Eq.~(\ref{nonnegative-sum-2}) can be written as 
\begin{equation}\label{trace-reservoir}
	\text{Tr}[\rho_\text{B}X^\dag(s)X]=\sum_{n,m} p_n e^{i(e_n^\text{B}-e_m^\text{B})s}|\bra{m}X\ket{n}|^2,
\end{equation}
where $e_{n,m}^\text{B}$ are the eigenenergies of the bath Hamiltonian $H_\text{B}$. Substituting Eq.~(\ref{trace-reservoir}) into Eq.~(\ref{nonnegative-sum-2}), we obtain
\begin{align}
	&\sum_{\alpha,\beta} v_\alpha^\ast \gamma_{\alpha\beta}(\omega) v_\beta \nonumber\\
	&\hspace{0.5cm}=\sum_{n,m} p_n |\bra{m}X\ket{n}|^2 
	\int_{-\infty}^{\infty}ds\, e^{i(\omega+e_n^\text{B}-e_m^\text{B}) s}\nonumber\\
	&\hspace{0.5cm}=2\pi\sum_{n,m} p_n |\bra{m}X\ket{n}|^2 \delta(\omega+e_n^\text{B}-e_m^\text{B}),
\end{align}
where $\delta(\cdot)$ denotes the Dirac delta function. The right-hand side is nonnegative (Q.E.D.).

Putting Eq.~(\ref{Gamma-decomposition}) into Eq.~(\ref{before-RWA}) and using RWA, we obtain the master equation in the following form: 
\begin{equation}\label{before-diagonalization}
	\frac{d}{dt}\rho_\text{S}(t)=-i\big[H_\text{LS},\rho_\text{S}(t)\big]+
	{\cal D}(\rho_\text{S}),
\end{equation}
where
\begin{equation}
	H_\text{LS}=\sum_{\omega}\sum_{\alpha,\beta}
	S_{\alpha\beta}(\omega)A^\dag_\alpha(\omega)A_\beta(\omega)
\end{equation}
is the so-called Lamb-shift Hamiltonian and the dissipator reads
\begin{align}
	&{\cal D}(\rho_\text{S})=\sum_\omega\sum_{\alpha,\beta}\gamma_{\alpha\beta}(\omega)
	\Big[A_\beta(\omega)\rho_\text{S}(t)A^\dag_\alpha(\omega)\nonumber\\
	&\hspace{3cm}
	-\frac{1}{2}\big\{A^\dag_\alpha(\omega)A_\beta(\omega),\rho_\text{S}(t)\big\}\Big].
\end{align}
Transferring back to the Schr\"odinger representation gives $H_\text{S}+H_\text{LS}$ instead of $H_\text{LS}$ in Eq.~(\ref{before-diagonalization}). The Lamb-shift Hamiltonian is small and usually can be neglected~\cite{Gardiner2010}. 

One obtains the usual GKSL form~(\ref{Lindblad}) of the master equation after the diagonalization of the matrix $\gamma_{\alpha\beta}(\omega)$. Here, the Lindblad operators are expressed through the system operators $A_\beta$ as follows (we assume the simplest case where the sum over $\omega$ contains only one term and the dependence of the operators on $\omega$ is not written explicitly): 
\begin{equation}
	L_k = \sum_\beta \sqrt{g_k}W_{k\beta}A_\beta.
\end{equation}
In this formula $g_k$ is an eigenvalue of the matrix $\gamma_{\alpha\beta}(\omega)$, and it is nonnegative because this matrix is positive semi-definite. The matrix $W$ diagonalizes the matrix $\gamma_{\alpha\beta}(\omega)$, i.e., $g=W\gamma W^\dag$, where $g$ is diagonal. 

If the system is a qubit, then the natural system operators $A_\alpha$ are the Pauli matrices $\sigma_+$, $\sigma_-$, $\sigma_z$.
In the simplest case $\gamma_{\alpha\beta}(\omega)\equiv\gamma\delta_{\alpha\beta}$, where $\gamma$ is a constant and $\delta_{\alpha\beta}$ is the Kronecker delta, one obtains the Lindblad operators proportional to $\sigma_-$ and $\sigma_z$. 
We note that the operator $\sigma_+$ is not relevant as it leads to the excitation of the qubit. 


\section{Derivation of the Gorini-Kossakowski-Sudarshan-Lindblad equation by using completely positive and trace preserving maps}
\label{Sec:AppendixGKSL-maps}
We now obtain the GKSL equation following Ref.~\cite{pearle2012}. For this, we start by listing the constraints necessary for the derivation of the GKSL equation:

(i)~the Markov constraint: the density matrix $\rho(t')\equiv\rho'$ at $t'>t$ depends on $\rho(t)\equiv\rho$, and not on density matrices at a \textit{range} of earlier times;

(ii)~the linearity constraint combined with the Markov constraint means the following:
\begin{equation}\label{linearity}
	\rho'_{ij}=\sum_{r,s=1}^N A_{ir,js}\rho_{rs},
\end{equation}
where $\rho'_{ij}$ and $\rho_{rs}$ are the matrix elements of $\rho'$ and $\rho$, respectively, defined in some orthonormal basis, and $A_{ir,js}$ can be the functions of $t, t'$;

(iii)~the Hermiticity constraint: $(\rho')^\dag=\rho'$;

(iv)~the trace constraint: $\text{Tr}\rho'=1$;

(v)~the positivity and complete positivity of $\rho'$.

It can be shown that the functions $A_{ir,js}$ satisfy the following condition:
\begin{equation}\label{hermiticity}
	A^\ast_{js,ir}=A_{ir,js}.
\end{equation}
Any Hermitian matrix can be written in terms of its eigenvalues $\lambda^\alpha$ and an orthonormal set of eigenvectors $\vec{E}^\alpha$. If $\rho$ is an $N \times N$ matrix, then $A$ is an $N^2 \times N^2$ matrix. For simplicity we consider $N=2$. A 4-component eigenvector $(\vec{E}^\alpha)^\intercal=(E^\alpha_{11}, E^\alpha_{12}, E^\alpha_{21}, E^\alpha_{22})$ can alternatively be presented as a $2\times 2$-matrix:
\begin{equation}\label{Ematrix}
	\hat{E}^\alpha=\left(\begin{array}{cccc}
		E^\alpha_{11} & E^\alpha_{12} \\
		E^\alpha_{21} & E^\alpha_{22} \\
	\end{array}\right).
\end{equation}
The matrix elements of the matrix $A$ read
\begin{equation}\label{Amatrix}
	A_{ir,js}=\sum_{\alpha=1}^4 \lambda^\alpha E^\alpha_{ir}(E^\alpha_{js})^\ast.
\end{equation}
Combining Eqs.~(\ref{linearity}) and (\ref{Amatrix}), one obtains
\begin{equation}\label{rho-prime-via-E}
	\rho'_{ij}=\sum_{\alpha=1}^4 \lambda^\alpha \left[\hat{E}^\alpha \rho (\hat{E}^\alpha)^\dag\right]_{ij}.
\end{equation}
Then, the trace constraint~(iv) can be written as
\begin{equation}\label{trace-constraint-v1}
	\sum_{\alpha=1}^4 \lambda^\alpha (\hat{E}^\alpha)^\dag \hat{E}^\alpha = \hat{1}.
\end{equation}
The condition for the orthonormality of the vectors $\vec{E}^\alpha$ reads
\begin{equation}\label{orthonormality}
	\text{Tr}[\hat{E}^\alpha (\hat{E}^\beta)^\dag]=\delta_{\alpha\beta}.
\end{equation}

The positivity of $\rho'$ means that $\forall \ket{v}$
\begin{equation}\label{positivity}
	\bra{v}\rho'\ket{v}\geq 0.
\end{equation}
The condition $\lambda^\alpha\geq 0$ $(\forall \alpha)$ is the sufficient condition of the positivity of $\rho'$, but it is not the necessary condition. 
The necessary condition for that is the complete positivity of $\rho'$, which can be defined as follows~\cite{Haroche2006}. Assume that at some moment of time in the past the system $S$ was entangled with another system $B$. Denote $\rho_{SB}$ the density matrix of the whole system. At present the system $B$ does not interact with $S$ and does not evolve. Then, the superoperator acting on $\rho_{SB}$ is ${\cal L}_S\otimes \hat{1}_B$, where ${\cal L}_S$ acts only on the system $S$ and describes its evolution. The complete positivity of ${\cal L}_S(\rho)$ means that ${\cal L}_S\otimes \hat{1}_B(\rho_{SB})$ is positive.

Equation~(\ref{rho-prime-via-E}) can be rewritten in terms of the Kraus operators
\begin{equation}\label{Kraus-operators}
	\hat{M}^\alpha=\sqrt{\lambda^\alpha}\hat{E}^\alpha
\end{equation}
as follows
\begin{equation}\label{rho-prime-via-M}
	\rho'=\sum_{\alpha=1}^4 \hat{M}^\alpha \rho (\hat{M}^\alpha)^\dag.
\end{equation}
In the notation~(\ref{Kraus-operators}) equation~(\ref{trace-constraint-v1}) becomes
\begin{equation}\label{trace-constraint-v2}
	\sum_{\alpha=1}^4 (\hat{M}^\alpha)^\dag \hat{M}^\alpha = \hat{1}.
\end{equation}

Considering Eq.~(\ref{rho-prime-via-E}) at $t'=t$, one can obtain the following formula:
\begin{equation}
	\delta_{ir}\text{Tr}\hat{E}^\beta=\lambda^\beta E^\beta_{ir}.
\end{equation}
Considering the case $\lambda^\beta\neq 0$, we obtain that the corresponding $\hat{E}^\beta$ is proportional to the identity matrix (there can be only one such eigenvector). The corresponding eigenvalue can be easily found. As a result,
\begin{equation}\label{N-squared}
	\hat{E}^{N^2}=\frac{1}{\sqrt{N}}\hat{1}, \qquad \lambda^{N^2}=N,
\end{equation}
while for $\alpha\neq N^2$ it holds
\begin{equation}
	\text{Tr}\hat{E}^\alpha = 0, \lambda^\alpha = 0.
\end{equation}
If the change of time is infinitesimal, $t'=t+dt$, then
\begin{align}
	&\lambda^{N^2}(dt)=N(1-c^{N^2}dt),\\
	&\hat{E}^{N^2}(dt)=\frac{1}{\sqrt{N}}(\hat{1}+\hat{B}dt),\\
	&\lambda^\alpha(dt)=c^\alpha dt \quad (\alpha\neq N^2),\\
	&\hat{E}^\alpha(dt)=\hat{K}^\alpha.
\end{align}
Then, Eq.~(\ref{rho-prime-via-E}) gives
\begin{align}
	&\frac{d\rho}{dt}=\frac{1}{2}[\hat{B}-\hat{B}^\dag,\rho]\\
	&-\frac{1}{2}\sum_{\alpha=1}^{N^2-1} c^\alpha 
	\left[(\hat{K}^\alpha)^\dag \hat{K}^\alpha \rho + 
	\rho (\hat{K}^\alpha)^\dag \hat{K}^\alpha -
	2\hat{K}^\alpha \rho (\hat{K}^\alpha)^\dag\right].
\end{align}
Denoting 
\begin{equation}
	-i\hat{H}=\frac{1}{2}(\hat{B}-\hat{B}^\dag)
\end{equation}
and
\begin{equation}
	\hat{L}^\alpha = \sqrt{c^\alpha}\hat{K}^\alpha,
\end{equation}
one obtains the GKSL equation~(\ref{Lindblad}). 


\section{Several notes on the Gorini-Kossakowski-Sudarshan-Lindblad equation for an adiabatically time-dependent Hamiltonian}\label{Sec:AppendixAdiabatic}

\subsection{Transfer to the interaction representation in the case of the time-dependent system Hamiltonian}

If the system Hamiltonian $H_\text{S}$ is time-dependent, the transfer to the interaction representation, instead of Eq.~(\ref{interaction-representation}), is given by the formula:
\begin{equation}
	O(t)={\cal U}^\dag(t,0)\,O\,{\cal U}(t,0),
\end{equation}
where the unitary evolution operator reads
\begin{equation}
	{\cal U}(t,0)={\cal T}_+ \text{exp}\left[-i\int_0^t d\tau H_\text{S}(\tau)\right]
\otimes e^{-iH_\text{B}t},
\end{equation}
with ${\cal T}_+$ denoting time ordering. If one further imposes a condition that the system Hamiltonian $H_\text{S}(t)$ changes with time adiabatically and makes similar steps~\cite{albash2012} to those ones given in Appendix~\ref{GSKL_Detail}, the master equation in the usual Lindblad form (but with time-dependent Lindblad operators) is obtained.

\subsection{The adiabatic conditions}

The adiabatic time dependence means that the following adiabatic condition is satisfied~\cite{albash2012}
	\begin{equation}\label{adiabatic-condition}
		\frac{g}{D_\text{min}^2 t_\text{f}}\ll 1,
	\end{equation}
where $t_\text{f}$ is the total evolution time of the system, $D_\text{min}$ is the minimum difference between the energies of the first excited and ground states of the system, and 
	\begin{equation}\label{parameter-h}
		g=\max_{s\in[0,1];a,b}|\bra{E_a(s)}\partial_s H_\text{S}(s)\ket{E_b(s)}|.
	\end{equation}
In Eq.~(\ref{parameter-h}) $s=t/t_\text{f}$ is the dimensionless time, $\ket{E_{a,b}(s)}$ are the instantaneous eigenvectors of the system Hamiltonian $H_\text{S}(s)$. 

For the Landau-Zener-St\"uckelberg-Majorana problem with the Hamiltonian 
	\begin{equation}
		H_\text{LZSM}(t)=\frac{1}{2}\Big(\Delta\,\sigma_x+v t\,\sigma_z\Big)
	\end{equation}
$D_\text{min}=\Delta$ and Eq.~(\ref{adiabatic-condition}) leads to the expected condition
	\begin{equation}
		\frac{\hbar v}{\Delta^2}\ll 1.
	\end{equation}
For a periodically driven qubit with the Hamiltonian~(\ref{qubit-Hamiltonian}) the estimates of the maximum in Eq.~(\ref{parameter-h}) lead to the two adiabatic conditions 
	\begin{equation}\label{adiabaticity-parameter}
		\frac{\hbar\omega A}{\Delta^2}\ll 1
	\end{equation}
and 
	\begin{equation}
		\frac{\hbar\omega A(|\varepsilon_0|+A)}{\Delta^3}\ll 1.
	\end{equation}
Assuming $|\varepsilon_0|+A<\Delta$, one can keep only the first one of these two conditions.

\subsection{Making the Hamiltonian time-independent}
	
We also note that for a periodically driven qubit with the Hamiltonian~(\ref{qubit-Hamiltonian}), considering the limiting case of small excitation amplitude, i.e., $A/\hbar\omega\ll 1$, and using the resonance approximation, the Hamiltonian can be made time-independent~\cite{Zvyagin1983}. 
At first, it is necessary to make a unitary transformation $\exp\left[i(A/2\hbar\omega)\sigma_z\sin\omega t\right]$ and keep in the resulting Hamiltonian the terms up to linear ones in the small parameter $A/\hbar\omega$. This gives
\begin{equation}
	H'(t)\approx H_0 - \frac{A}{\hbar\omega}\frac{\Delta}{2}\sin\omega t\sigma_y,
\end{equation}
where the static Hamiltonian $H_0$ is defined in Eq.~(\ref{H_0}). Next, one transfers to the static eigenstates basis and keeps only the resonant terms. 
This gives
\begin{equation}
	H''(t)\approx -\frac{1}{2}\sqrt{\Delta^2+\varepsilon_0^2}\,\sigma_z + \frac{A}{\hbar\omega}\frac{\Delta}{4}\left(e^{i\omega t}\sigma_+ 
	+e^{-i\omega t}\sigma_-\right).
\end{equation}
Then, the next unitary transformation $\exp[i\omega t\sigma_z/2]$ leads to the following time-independent Hamiltonian
\begin{equation}
	\tilde{H}\approx -\frac{1}{2}\left(\sqrt{\Delta^2+\varepsilon_0^2}-\hbar\omega\right)\sigma_z
	+\frac{A}{\hbar\omega}\frac{\Delta}{2}\sigma_x.
\end{equation}
An analogous procedure of making the Hamiltonian time-independent can be done for a multilevel system, and in some cases this allows the analytical solution to the GKSL equation to be obtained~\cite{Zvyagin2020}.


\section{Transfer from the instantaneous to the first superadiabatic basis for a periodically driven qubit}\label{Appendix-superadiabatic}
The transfer matrix from the instantaneous to the first superadiabatic basis is constructed from the eigenvectors of the Hamiltonian~(\ref{Hamiltonian-adiabatic}).
Keeping the terms up to the first order of $a$, we obtain the following eigenvectors of the matrix in Eq.~(\ref{Hamiltonian-adiabatic}):
\begin{equation}\label{superadiabatic-eigenvector-1}
	\left(\begin{array}{cccc}
		x_{-} \\
		y_{-} \\
	\end{array}\right)\approx
	\left(\begin{array}{cccc}
		i \\
		B(t) \\
	\end{array}\right)e^{i\varphi_{-}(t)},
\end{equation}
\begin{equation}\label{superadiabatic-eigenvector-2}
	\left(\begin{array}{cccc}
		x_{+} \\
		y_{+} \\
	\end{array}\right)\approx
	\left(\begin{array}{cccc}
		-i B(t) \\
		1 \\
	\end{array}\right)e^{i\varphi_{+}(t)},
\end{equation}
with 
\begin{equation}\label{B}
	B(t)=-\hbar \dot{\theta}(t)/2\Delta E(t),
\end{equation}
which is of the order of $a$ [see Eq.~(\ref{nonadiabatic-coupling-SectionD})].
The phase factors $\varphi_{\pm}(t)$ can be obtained from the parallel transport condition~\cite{simon1983}, which reads
\begin{equation}
	\left(x_{\pm}^\ast,  y_{\pm}^\ast\right)\cdot
	\left(\begin{array}{cccc}
		\dot{x}_{\pm} \\
		\dot{y}_{\pm} \\
	\end{array}\right)=0.
\end{equation}
This condition leads to constant $\varphi_{+}$ and $\varphi_{-}$. Taking $\varphi_{+}=0$, $\varphi_{-}=-\pi/2$, and keeping terms up to the first order in $a$, we obtain the following transfer matrix:
\begin{equation}
	U_1(t)\approx
	\left(\begin{array}{cccc}
		1 & -i B(t) \\
		-i B(t) & 1 \\
	\end{array}\right).
\end{equation}
The Schr\"odinger equation in the first superadiabatic basis reads
\begin{equation}\label{Schrodinger-1-superadiabatic-basis}
	i\hbar \frac{d\ket{\psi_1(t)}}{dt}=H_1(t)\ket{\psi_1(t)},
\end{equation}
where
\begin{align}\label{Hamiltonian-1-superadiabatic}
	&H_1(t)=
	U_1^{-1}(t)H_\text{ad}(t)
	U_1(t)
	-i\hbar U_1^{-1}(t)
	\dot{U}_1(t)\nonumber\\
	& \hspace{1cm}=-\left(\frac{\Delta E(t)}{2}+\frac{\hbar^2\dot{\theta}^2(t)}{4\Delta E(t)}\right)\sigma_z - \dot{B}(t) \sigma_x.
\end{align} 
As the differentiation of the function $B(t)$, Eq.~(\ref{B}), and simple estimates show, the off-diagonal terms of the matrix in Eq.~(\ref{Hamiltonian-1-superadiabatic}) are of the order of $a^2$. The eigenvalues differ from the eigenvalues of the instantaneous Hamiltonian by the terms of the order of $a^2$ [see Eqs.~(\ref{nonadiabatic-coupling-SectionD}, \ref{parameter-a})].

\bibliography{ReferencesJabRef}
	
\end{document}